# A highly sensitive, self-adhesive, biocompatible DLP 3D printed organohydrogel for flexible sensors and wearable devices

Ze ZHANG[a, §], Kewei SONG[a, §], Kayo HIROSE [b, §], Jianxian HE[a], Qianhao LI[c], Yannan LI[d], Yifan PAN[a], Mohamed ADEL[a,e], Rongyi ZHUANG[a], Shogo IWAI[d], Ahmed M. R. FATH EL-BAB[e], Hui FANG[f,*], Zhouyuan YANG[c,*], Shinjiro UMEZU[a,*]

[a]*Graduate School of Creative Science and Engineering, Department of Modern Mechanical Engineering, Waseda University, 3-4-1 Okubo, Shinjuku-ku, Tokyo 169-8555, Japan.*

[b]*Anesthesiology and Pain Relief Center, The University of Tokyo Hospital, 7-3-1 Hongo, Bunkyo-ku, Tokyo 113-8655, Japan.*

[c]*West China School of Medicine, Sichuan University, Chengdu, Sichuan 610064, China.*

[d]*Graduate School of Advanced Science and Engineering, Department of Integrative Bioscience and Biomedical Engineering, Waseda University, 3-4-1 Okubo, Shinjuku-ku, Tokyo 169-8555, Japan.*

[e]*Department of Mechatronics and Robotics Engineering, Egypt-Japan University of Science and Technology (E-JUST), Alexandria 21934, Egypt*

[f]*Department of Mechanical Engineering, Sichuan University, Chengdu, Sichuan 610064, China.*

*§: The authors contributed equally.*

**\*Corresponding authors:**

*Shinjiro UMEZU, Ph.D., Professor, E-mail: umeshin@waseda.jp*

*Hui FANG, Ph.D., Professor, E-mail: jfh@scu.edu.cn*

*Zhouyuan YANG, Ph.D., E-mail: zhouge314@163.com*

**Abstract**

With the growing demand for personalized health monitoring, wearable sensors have gained significant attention in medical diagnostics and physiological monitoring. Hydrogels, due to their mechanical properties and structural similarity to biological tissues, are ideal flexible sensing materials. However, existing hydrogels face challenges in stability, biocompatibility, adhesion, and long-term comfort, particularly in dynamic applications. This study introduces a highly sensitive, self-adhesive, and biocompatible organohydrogel fabricated via digital light processing (DLP) 3D printing. By integrating an entanglement-dominated crosslinking mechanism with chemical and physical crosslinking strategies, the hydrogel achieves superior elasticity, mechanical strength, and durability. Methacrylic anhydride-grafted κ-carrageenan (MA-κ-CA) serves as the primary hydrogel network, with optimized grafting rates to enhance tensile properties and strain modulation. The copolymer network of MA-κ-CA and ACMO (4-acryloylmorpholine) benefits from steric hindrance effects, improving swelling integrity and long-term stability. Experimental results demonstrate sustained adhesion and structural integrity under prolonged skin exposure, making the hydrogel suitable for extended wear. The entanglement-enhanced hydrogel exhibits excellent tensile resilience, flexibility, and strain-sensing performance. In vitro studies confirm its biocompatibility. Furthermore, the hydrogel shows promise in wearable smart devices, particularly for cervical spine monitoring and sports rehabilitation. A convolutional neural network (CNN)-based system enables real-time multi-channel analysis of cervical motion, validating its potential as a high-sensitivity flexible sensor for health monitoring and injury prevention. The proposed DLP 3D-printed hydrogel offers significant applications in flexible electronics, wearable sensors, and biomedical fields, presenting new opportunities for advanced health-monitoring technologies.

# 1. Introduction

Wearable sensors are pivotal in advancing personalized healthcare, clinical diagnostics, and physiological monitoring[1, 2]. Hydrogels, characterized as three-dimensional elastic polymer networks with high water content, have gained substantial attention for their remarkable mechanical and structural similarity to biological tissues. A distinct advantage of hydrogels lies in their tunable network structure, electrochemical properties, mechanical performance, and biological functionality, allowing for customization across a wide range of applications, particularly in the development of flexible ionic devices. Despite their potential, current research places stringent demands on hydrogels, necessitating outstanding stability, biocompatibility, adhesion, and long-term comfort, all while maintaining reliable performance under dynamic conditions[3]. These multifaceted functional requirements not only expand the application of hydrogels in biomedicine and sensor technologies but also present considerable challenges in their design and manufacturing. Addressing these challenges requires innovative approaches to materials engineering and advanced manufacturing techniques, particularly in additive manufacturing[4].

To enhance the mechanical performance and stability of hydrogels, researchers have employed various crosslinking strategies. Copolymer hydrogels are synthesized by polymerizing multiple monomers, allowing tunable properties. The combination of hydrophilic and hydrophobic monomers enhances mechanical strength, water absorption, and thermal stability, making them suitable for drug delivery, wound dressings, and flexible scaffolds. However, challenges remain in synthesis reproducibility and stability under extreme conditions[5]. Dual-crosslinked hydrogels integrate physical and chemical crosslinking, where non-covalent interactions (hydrogen bonds, electrostatic forces, ionic bonds) complement covalent bonds to improve strength, toughness, and self-healing while maintaining flexibility. This design is widely used in soft electronics, sensors, and tissue engineering, but its complex synthesis process requires precise control to prevent excessive crosslinking-induced brittleness[6, 7]. Ionic crosslinked hydrogels utilize electrostatic interactions between metal ions (such as $Ca^{2+}$, $Na^+$) and polymer chains, offering tunability and low toxicity, making them ideal for biomedical applications. They are easy to prepare, free from toxic crosslinkers, and stable in aqueous environments, but their relatively weak crosslinking strength limits resistance to high stress[8, 9].

With increasing demands for the performance of hydrogels, traditional crosslinking methods have shown limitations in certain applications. Particularly in cases where high mechanical performance and good flexibility are required simultaneously, conventional crosslinking designs may not meet the needs. Therefore, the introduction of the entanglement-dominated over crosslinking mechanism has emerged as an innovative design strategy[10, 11]. In this design, polymer chains form a crosslinked network primarily through high-density entanglements, rather than relying entirely on chemical crosslinking. By increasing the degree of chain entanglement, the hydrogel can achieve more flexible mechanical properties while maintaining high strength[12]. This mechanism is especially beneficial for applications requiring dynamic responsiveness and excellent fatigue performance, such as motion monitoring and flexible sensors[13]. This strategy provides new insight into hydrogel design, addressing the brittleness issues inherent in traditional crosslinking methods and expanding their potential applications in biomedical and smart sensor fields.

In recent years, natural-origin polymers have attracted widespread attention due to their similarity to the extracellular matrix (ECM), high chemical multifunctionality, controllable biodegradability, and good biocompatibility[14]. In hydrogel design, natural polymers are widely used to improve the biological properties, mechanical properties, and degradation control of hydrogels[15]. Natural polymers typically exhibit good biocompatibility and degradability, making them compatible with the biological environment, and they can degrade into non-toxic metabolic products in vivo or in vitro, avoiding the potential toxicity issues associated with synthetic polymers[16]. Additionally, natural polymers can provide a physical and chemical environment similar to the ECM, which is crucial for cellular adhesion, proliferation, and differentiation[15]. These advantages make them indispensable materials in the biomedical field, particularly in tissue engineering, drug delivery, and cell culture applications.

Natural polymers can form hydrogels through various crosslinking methods, enhancing their tunability [17]. Kappa-carrageenan, a widely used natural polysaccharide with excellent biocompatibility and biodegradability, forms ionic crosslinked networks through interactions with calcium ions, making it a promising candidate for flexible hydrogel fabrication. Its physical crosslinking properties facilitate the formation of a flexible network that supports cell adhesion and controlled biodegradability; however, its low mechanical strength limits its applicability in high-stress environments[17-19]. Alginate forms stable ionic crosslinked hydrogels upon interaction with divalent metal ions and exhibits superior biocompatibility, making it suitable for flexible sensors and bioelectronic interfaces. However, its mechanical strength remains insufficient under high-stress conditions, necessitating copolymerization or composite strategies to enhance its structural stability[20]. Gelatin, commonly crosslinked via thermal or photopolymerization, possesses excellent cell adhesion and biodegradability, making it a viable candidate for bioelectronics and flexible sensing applications. However, its poor thermal stability leads to structural degradation in humid or high-temperature environments, necessitating hybridization with other polymers for improved mechanical performance[21, 22]. Hyaluronic acid, known for its high hydrophilicity and flexible network structure, has been extensively explored in bioelectronic sensors. Its chemically or photochemically crosslinked hydrogel networks promote cell migration and enhance the biocompatibility of flexible electronic devices. Nevertheless, its low mechanical strength requires reinforcement through polymer blending[23, 24]. Chitosan, which forms hydrogels via chemical or ionic crosslinking, is widely used in flexible sensors and wearable electronics. However, its pH-dependent solubility and post-crosslinking brittleness restrict its applications, often necessitating composite formulations with alginate or gelatin to optimize mechanical properties and environmental stability[25, 26]. Although natural polymers hold great potential for flexible hydrogel-based sensors, challenges related to mechanical robustness and long-term stability remain. Current research focuses on hybrid strategies combining natural and synthetic polymers to enhance material performance, expanding their applications in flexible electronics and wearable sensor technology[15]. This study presents a highly sensitive, self-adhesive, and biocompatible organohydrogel fabricated using digital light processing (DLP) 3D printing technology. By introducing an entanglement-dominated crosslinking mechanism combined with both chemical and physical crosslinking strategies, the hydrogel exhibits excellent mechanical strength and durability while maintaining high elasticity and flexibility. Methacrylic anhydride-grafted κ-carrageenan (MA-κ-CA) was utilized as the primary hydrogel network, and the crosslinking density was optimized by adjusting the grafting degree of methacrylic anhydride to enhance its tensile properties and strain modulation capability. Additionally, in the MA-κ-CA and ACMO (4-acryloylmorpholine) copolymer network, the steric hindrance effect improves the swelling integrity of the hydrogel and enhances its long-term stability. In conclusion, the proposed DLP 3D-printed hydrogel offers significant applications in flexible electronics, wearable sensors, and biomedical fields, presenting new opportunities for advanced health-monitoring technologies.

## 2. Method and experiment

### 2.1 Materials

All reagents were used as received without further purification. 4-Acryloylmorpholine (ACMO), glycerol (molecular biology, ≥99 %), κ-Carrageenan (κ-CA), methacrylic anhydride (MA, ≥94 %), Calcium chloride($CaCl_2$, ≥97 %), potassium Chloride(KCl), sodium chloride(NaCl, ≥99 % ), and sodium hydroxide(NaOH, ≥97 %) were all bought from Sigma–Aldrich Co., Ltd. Deionized water (DI water) was used to configure the required solutions.

### 2.2 Synthesis of Methacrylic anhydride-κ-Carrageenan

Methacrylic anhydride-κ-Carrageenan (MA-κ-CA) was synthesized as described previously[27]. Briefly, κ-CA was dissolved in deionized (DI) water with 1 % (w/v) at 50 °C. Then MA was added to this solution and react for 6 hours at 50 °C. During the reaction, pH (8.0) was periodically adjusted with 5.0 M NaOH solution in DI water.  After reaction, the solution was transferred into a dialysis membrane (typical molecular weight cut-off 14,000 Da) for 3 days at 4 °C. The dialysis liquid (DI water) was exchanged every 12 h to remove unreacted MA. Purified MA-κ-CA solutions were frozen in liquid nitrogen and then dried in in a freeze-dryer to obtain the white powder. The obtained polymer samples were stored at -20 °C and protected from light until further use. The methacrylated degree (DM) was obtained by adding different volumes of MA (4% (v/v), 8 % (v/v) and 12 % (v/v)).

### 2.3 Preparation of hydrogel precursor

First, calcium chloride (0.02 M or 0.1 M in DI water) was added to DI water as a cross-linking agent and conductive ion. After thorough stirring for dissolution, MA-κ-CA (10 wt% in DI) was added, and the solution was heated to 50 °C. After mixing, the solution was cooled to room temperature to obtain the MA-κ-CA hydrogel precursor solution. ACMO (70 wt%), glycerol (20 wt%), and MA-κ-CA hydrogel solution (10 wt%) were added to this hydrogel precursor solution, along with TPO (1.5 wt% of ACMO) as a photoinitiator. The mixture was stirred thoroughly under light shielding to obtain the photopolymerizable hydrogel precursor solution.

### 2.4 3D print

The 3D printing process for the hydrogel precursor was conducted using a commercial DLP 3D printer (Anycubic Photon D2). The 3D models of the hydrogel were designed in SolidWorks (Dassault Systèmes S.A.) and sliced using Photon Workshop from Anycubic. The layer thickness was set to 50 μm, and the wavelength of the UV laser used for printing was 405 nm. Each layer was cured for 6 seconds, during which the hydrogel precursor underwent photopolymerization. After printing, the samples were quickly washed with water to remove unreacted precursor materials.

### 2.5 Characterizations

#### 2.5.1 Characterization of MA-κ-CA

The DM was determined using 1H **nuclear magnetic resonance spectroscopy (NMR)**. κ-CA and MA-κ-CA was dissolved in deuterated water ($D_2O$) and their 1H NMR spectra were obtained at 50 ℃ at a frequency of 600 MHz on a AVANCE 600 NEO(Bruker Corporation, USA). The obtained chemical shifts were normalized against the protons of the methylene group of the D-galactose units (G) as an internal standard[27], which is present at 3.89-4.00 ppm. The peak area of acrylic protons of the methacrylated compound appeared at 5.5-6.0 ppm. The peak area of methyl proton signals appeared at around 1.9-2.0 ppm. The DM was calculated referring to the peaks at 5.5-6 ppm (acrylic protons) and 1.9-2 ppm (methyl protons) as percentage (%) of the free hydroxyl groups (–OH) substituted with methacrylate groups[27]. **FTIR-ATR** analysis was performed using an FT/IR-4200 spectrometer (JASCO, Japan) with a resolution of 4 cm⁻¹ and an average of 16 scans. For sample preparation, approximately 80 mg of potassium bromide (KBr) was used as the matrix, mixed with 0.1 mg of MA-κ-CA.

**2.5.2 Swelling behavior**

The influence of the degree of methacrylation and the crosslinking mechanism on the swelling behavior and stability of hydrogels was determined by evaluating hydration kinetics and dissolution behavior.

**2.5.3 Scanning electron microscopy of dried hydrogels:**

The microstructure of the hydrogels was evaluated using a Scanning Electron Microscope (SEM, SU-8240, Hitachi High-Tech Corporation, Japan) at an acceleration voltage of 10 kV. Initially, the hydrogel samples (swollen in DI water for 40 minutes) were immersed in liquid nitrogen slush, transferred to centrifuge tubes, and freeze-dried for 24 hours. The dried samples were fractured and mounted on sample stages using double-sided carbon tape. The cross-sections were coated with a platinum layer using a MC1000 Ion Sputter (Hitachi High-Tech Corporation, Japan). The quantification of pore size distribution was conducted using ImageJ for statistical analysis.

**2.5.4 Tensile and adhesion testing**

Tensile tests were conducted using an MCT-2150 bench-top tensile testing machine (A&D). Dog-bone-shaped tensile specimens were printed according to the ISO 527-4 standard. The tests were performed at speeds of 10, 40, and 100 mm/min, corresponding to a strain rate of 0.0067 s⁻¹ (Figure S 2.5.1, Supporting Information). The Young's modulus was calculated from the first 30% of strain. Additionally, 180° peeling tests were performed using the same setup (**Figure S 2.5.2, Supporting Information**).

**2.5.5 Conductivity measurement**

The variation of electrical resistance was calculated by using the equation (1).

$$\frac{\Delta R}{R_0} = \frac{R - R_0}{R_0} \tag{1}$$

where $R_0$ is the resistance without strain and R denotes the real time resistance.

Other experimental tests include the test of water contact angle, porosity, piezoelectric coefficient, please refer to the supplementary material for specific methods.

**2.5.6 Cytocompatibility**

L929 cells were used to examine cell viability of the scaffolds. L929 cells were purchased from the American Type Culture Collection (ATCC, USA) and cultured in Dulbecco's Modified Eagle Medium (DMEM, #11965092, Gibco, USA). The medium was supplemented with 10% fetal bovine serum (#10100147C, Gibco, USA) and 1% penicillin-streptomycin solution (#SV30010, HyClone, USA). The cells were seeded on the surface of materials in 24-well plates at a density of 4*10⁴ cells/well and then cultured at 37°C in a CO₂ incubator (#3140, Thermo Scientific, USA). To assess L929 cells' viability after co-cultured with the scaffolds for 2 days, the survival rate was measured using the Calcein/PI Cell Viability Assay Kit (#C2015, Beyotime, China) according to its manufacturer's protocol. Live cells (stained green by calcein with excitation at 494 nm) and dead cells (stained red by PI with excitation at 535 nm) were observed using a confocal laser scanning microscope (#LSM 980, ZEISS, Germany). Three parallel samples were designed for random image capture and cells were counted automatically using the ImageJ software (1.53K version, National Institutes of Health, USA).

**2.5.7 Statistical Analysis**

All tests were repeated three times and data were represented as the mean ± standard deviation (SD).

## 3. Result and Discussion

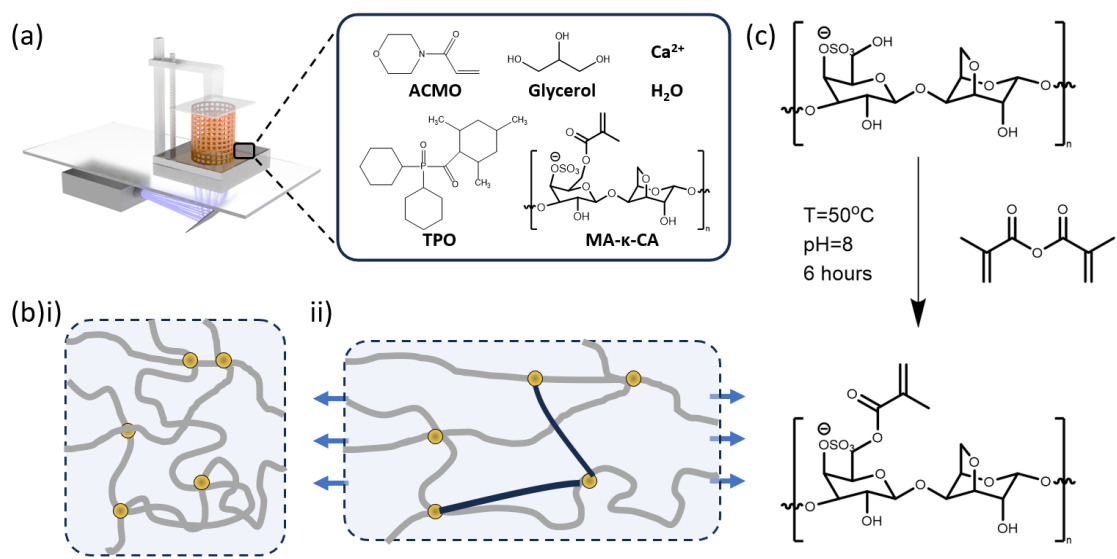

**Figure 1. Overview of the proposed hydrogel design.** (a) Schematic of the hydrogel stereolithographic formation based on DLP 3D printing, (b) Tensile force transmission process of the entanglement-dominated hydrogel under stress, (c) Molecular reaction scheme.

The hydrogel design proposed in this study combines both chemical and physical crosslinking strategies to achieve a material with excellent mechanical properties, suitable for applications such as motion monitoring. The hydrogel is primarily designed using a DLP 3D printing technique, which ensures precise control over the material's geometry and allows for the construction of three-dimensional structures with high resolution. The hydrogel precursor solution is prepared by dissolving MA-κ-CA in deionized water, followed by the addition of calcium chloride, glycerol, and the photoinitiator TPO. The mixture is then photopolymerized layer by layer using a 3D printer equipped with a DLP system. This allows the hydrogel to undergo a controlled photopolymerization process, where each layer is exposed to UV light, curing it and forming the final structure. This technique is ideal for fabricating hydrogels with high spatial resolution and tunable mechanical properties, as it allows precise control over the material's crosslinking density and geometry.

In the hydrogel network, the combination of entanglement and crosslinking plays a crucial role in determining its mechanical properties. As illustrated in **Figure 1 (b)**, the hydrogel structure is dominated by entanglements, which are more prevalent than crosslinks. This results in a network where the chains are loosely connected, providing the material with the flexibility to undergo deformation under stress. This characteristic is particularly important for applications where the hydrogel needs to accommodate movement, such as in motion monitoring systems. The entangled network allows for efficient stress transfer across the material, enhancing its toughness and durability.

**Figure 1 (c)** shows the molecular reaction scheme for the synthesis of MA-κ-CA. The reaction involves the substitution of hydroxyl groups on the κ-carrageenan backbone with methacrylate groups, facilitated by methacrylic anhydride. The reaction is carried out at 50°C for six hours, with the pH adjusted periodically to maintain optimal conditions for the substitution reaction. The degree of methacrylation (DM) can be controlled by varying the concentration of methacrylic anhydride during the synthesis process, which in turn influences the hydrogel's swelling behavior, mechanical properties, and crosslinking density.

The resulting hydrogel exhibits a unique combination of physical and chemical crosslinking. The physical crosslinking arises from the entanglements of polymer chains, while the chemical crosslinking is induced by the covalent bonds formed between the methacrylate groups during photopolymerization. The balance between these two types of crosslinking contributes to the hydrogel's remarkable mechanical properties, such as high elasticity and resistance to mechanical failure under strain. The entanglement-dominated structure provides flexibility, while the chemical crosslinks ensure stability and resistance to deformation. This dual crosslinking strategy makes the hydrogel an ideal candidate for motion monitoring applications, where flexibility, durability, and sensitivity to mechanical changes are crucial. The hydrogel can easily adapt to different deformation patterns, such as stretching, compressing, or bending, without compromising its structural integrity. Additionally, its responsiveness to mechanical stimuli enables the development of highly sensitive sensors for real-time monitoring of body movements or other dynamic processes.

In summary, the proposed hydrogel design, which combines DLP 3D printing with a dual crosslinking strategy, provides a material with excellent mechanical properties that can meet the demands of motion monitoring applications. The balance between entanglement and crosslinking in the hydrogel's network ensures both flexibility and stability, making it suitable for use in dynamic environments where durability and high performance are essential.

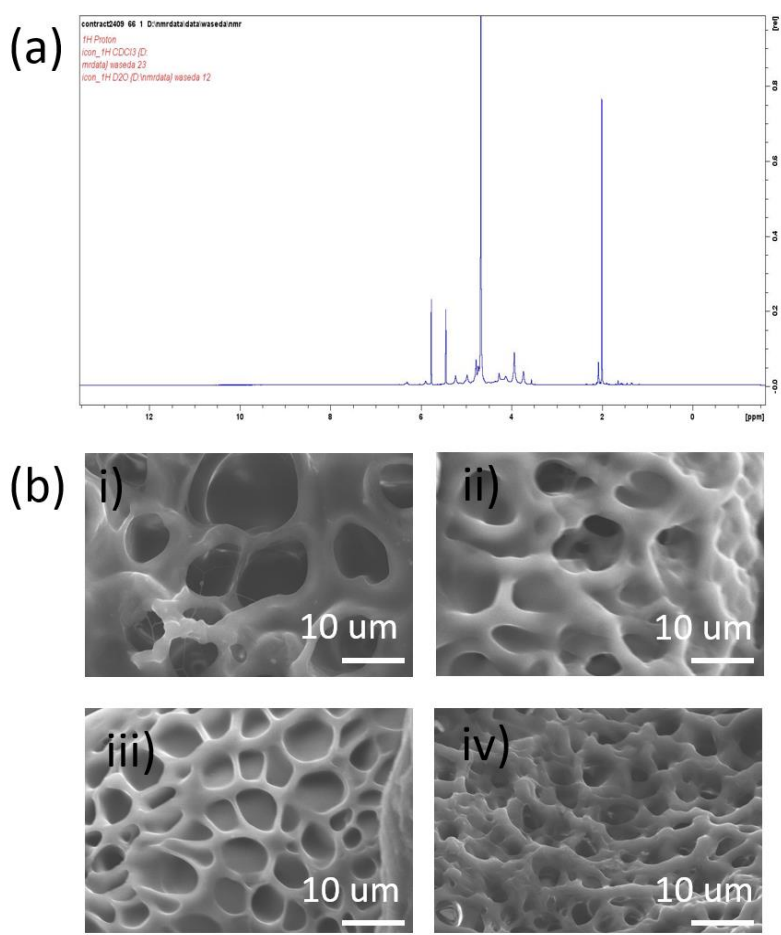

**Figure 2. Hydrogel characterization.** (a) NMR test for CA grafting rate, (b) Hydrogel microstructure, i)-iv) grafting rates of 0 %, 8 %, 12 %, 16 %.

Methacrylated-κ-carrageenan (MA-κ-CA) with different degrees of methacrylate (DM) were synthesized by substituting hydroxyl groups on κ-CA with methacrylate groups. Proton nuclear magnetic resonance ($^1$H NMR) spectroscopy confirmed the methacrylation of κ-CA, with the appearance of a double peak of vinyl (δ=5.5-6 ppm) and a peak of the methyl group (-CH$_3$, δ=1.9-2 ppm) belong to methyl methacrylate group. To quantify the DM, substitution degree of the free hydroxyl group on backbone of κ-CA was assessed by comparing the average integrated intensity of the methyl proton peaks of methyl methacrylate group with the methylidene group in the β-D galactose (G6) of κ-CA (δ=3.9 ppm). The addition of 4%, 8%, 12% and 16% (v/v) of MA to κ-CA during the synthesis process allowed the preparation of different MA-κ-CA polymers with DMs of 4%, 8%, 12% and 37.26%, respectively.

To further analyze the microstructure of the hydrogel, scanning electron microscopy (SEM) was used to observe the morphology of the hydrogel samples with different grafting rates. **Figure 2 (b)** shows the microstructure of the hydrogels with grafting rates of 0%, 8%, 12%, and 16%. **Figure 2 (b) i)** presents the hydrogel with a 0% grafting rate, which has larger, unevenly distributed pores with loosely connected pore structures, exhibiting low crosslinking density and poor mechanical strength. This suggests that the un-grafted hydrogel structure is relatively loose and suitable for applications where high porosity is required but mechanical performance is not critical. **Figure 2 (b) ii)** shows the hydrogel with an 8% grafting rate, where the pore morphology has improved, with more uniform pore sizes and tighter connections between the pores. This indicates that as the grafting degree increases, the mechanical strength and stability of the hydrogel improve. **Figure 2 (b) iii)** shows the hydrogel with a 12% grafting rate, where the pore structure is more regular, and the hydrogel's mechanical properties and stability are further enhanced. The higher grafting degree helps maintain high hydration and permeability of the hydrogel's pores, making it suitable for applications such as cell culture or drug delivery. **Figure 2b iv)** shows the hydrogel with a 16% grafting rate, where the pore structure is tighter, with smaller pore sizes, and the overall structure appears more robust. This suggests that hydrogels with higher grafting rates exhibit stronger mechanical strength and better stability, making them suitable for applications requiring high mechanical strength and stability, such as soft scaffolds and biosensors. From the SEM images, it is evident that as the degree of methacrylation increases, the hydrogel's microstructure becomes denser and more uniform, with optimized pore distribution. This indicates that increasing the grafting degree helps enhance the hydrogel's mechanical properties and structural stability, providing flexible design options for its applications in tissue engineering, wound dressings, and biosensors in the biomedical field.

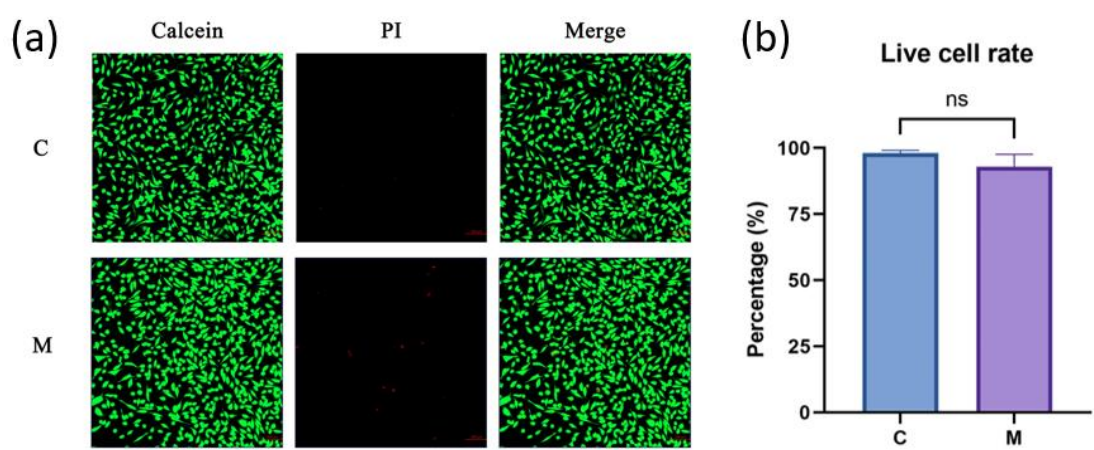

**Figure 3. Hydrogel biocompatibility test results.** (a) Live and dead cell staining, (b) Cell counting statistics results.

To evaluate the biocompatibility of the prepared hydrogel, L929 mouse fibroblast cells were used for cell viability testing. The cells were cultured in Dulbecco's Modified Eagle Medium (DMEM) supplemented with 10% fetal bovine serum and 1% penicillin-streptomycin solution. The cells were seeded at a density of 4 × 10$^4$ cells/well in 24-well plates and cultured at 37°C in a CO$_2$ incubator.

After co-culturing with the hydrogel for 2 days, cell viability was assessed using the Calcein/PI Cell Viability Assay Kit. Live cells were stained green with Calcein (excited at 494 nm), while dead cells were stained red with PI (excited at 535 nm). Fluorescence microscopy was used to observe and analyze the live and dead cells to further assess the biocompatibility of the hydrogel.

As shown in **Figure 3**, the Calcein/PI staining results indicated that both the control group (C) and the material group (M) had almost no significant difference in cell viability. In both groups, the majority of cells were green, indicating that most of the cells were viable. The number of red PI-stained dead cells was minimal, further confirming the excellent biocompatibility of the hydrogel with L929 cells. Quantitative analysis of cell viability showed that the cell viability in both groups was close to 100%, with no significant difference ($p > 0.05$), indicating that the hydrogel did not exhibit any toxicity or inhibitory effect on the cells. In conclusion, the proposed hydrogel demonstrated excellent biocompatibility in L929 cell cultures, with no adverse effects on cell growth and survival. This indicates that the hydrogel is suitable for biomedical applications, such as tissue engineering and wound dressings.

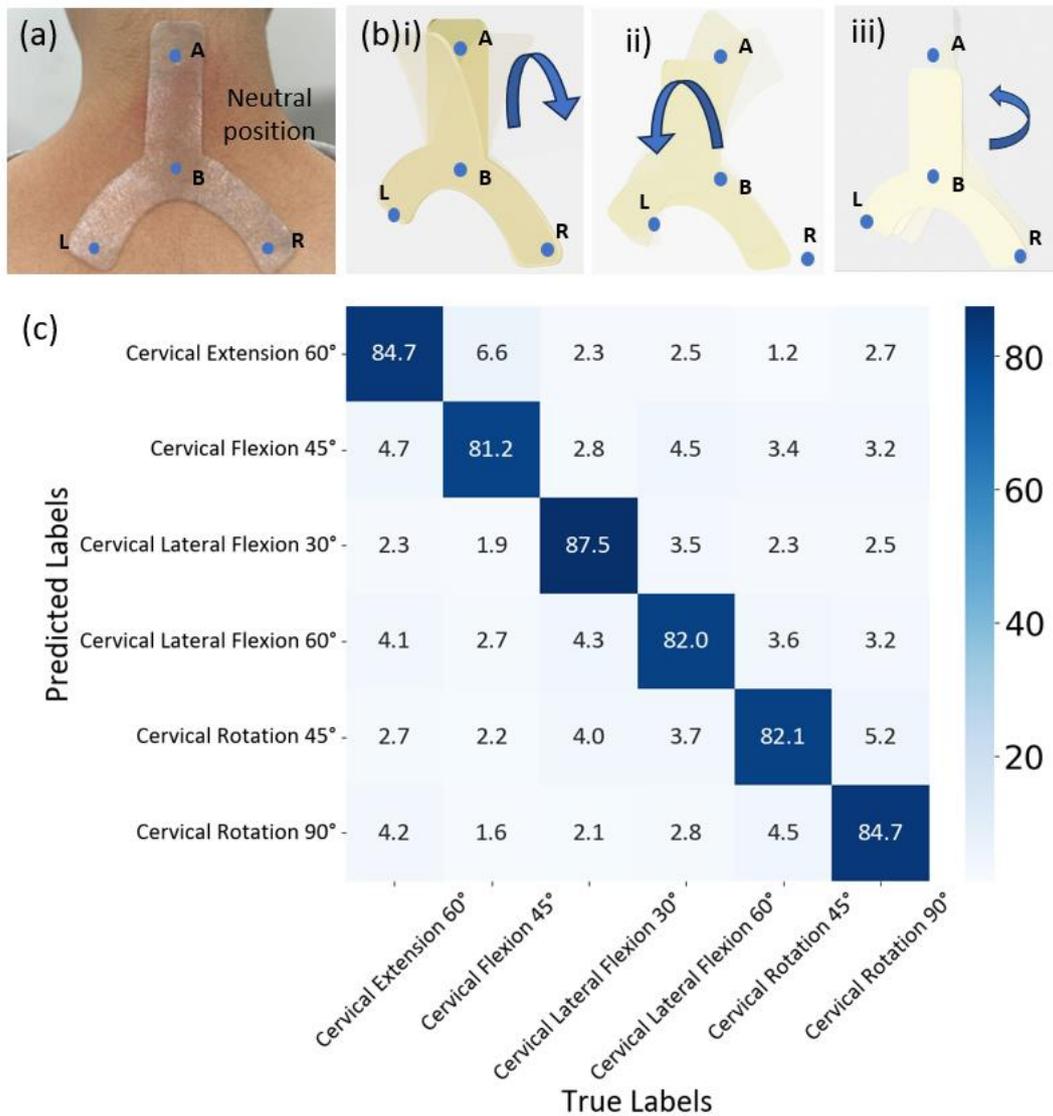

**Figure 4. Multi-channel neck condition monitoring.** (a) Hydrogel arrangement diagram, (b) Schematic diagram of neck activity modes, i) back and forth activity, ii) side head, and iii) head turn, (c) State analysis of neck activity in multi-channel signaling state based on CNN model.

**Figure 4** presents the multi-channel signal-based neck condition monitoring system and its analysis results. The system utilizes resistance change sensors or similar multi-channel signal technologies, combined with a convolutional neural network (CNN) model (please refer to supplementary materials for model details), to perform real-time monitoring and classification of neck motion states.

**Figure 4 (a)** shows the arrangement diagram of the hydrogel. The diagram highlights four key points (A, B, L, R), which are positioned at different areas of the neck. During the monitoring process, the changes in resistance or other signal characteristics at these points are recorded to reflect different neck motion states. The key to the hydrogel arrangement lies in accurately capturing changes in motion states and collecting data through various sensor signals. **Figure 4(b)** illustrates three different neck motion activity modes. **Figure 4(b) i)** represents forward and backward movements, flexion and extension of the neck, typically characterized by the head moving back and forth, which mainly tests the neck's extension and flexion abilities. **Figure 4(b) ii)** depicts side head movements, lateral flexion, which primarily tests the neck's lateral bending ability. **Figure 4(b) iii)** shows neck rotation, head turning, which primarily tests the neck's rotation capability. These three common neck motion modes are used to assess the variations in sensor signals from the hydrogel under different motion states.

**Figure 4 (c)** shows the analysis of multi-channel signal states for neck activity based on the CNN model. Each value in the matrix represents the model's prediction accuracy for different motion states. The x-axis and y-axis represent the predicted labels and true labels, respectively, and the values in the matrix represent the classification accuracy of the model for various activity modes. Overall, the model demonstrates high accuracy across most activity modes, particularly for common neck motions such as Cervical Extension 60° and Cervical Rotation 90°, where the prediction accuracy exceeds 80%. These results indicate that the CNN model based on multi-channel signals can effectively recognize and classify different neck motion states. This monitoring system enables high-precision real-time tracking of neck motion states, providing reliable data support for clinical diagnosis, physical rehabilitation, and daily health management.

## 4. Conclusion

This study introduces a highly sensitive, self-adhesive, and biocompatible organohydrogel fabricated via DLP 3D printing, designed for flexible sensing and wearable applications. By integrating an entanglement-dominated crosslinking mechanism with chemical and physical crosslinking strategies, the hydrogel achieves superior tensile strength, flexibility, and durability, while maintaining excellent biocompatibility. The incorporation of methacrylic anhydride-grafted κ-carrageenan (MA-κ-CA) optimizes the network structure, enhancing strain modulation and swelling integrity, thereby improving its long-term stability. Experimental results demonstrate that the hydrogel maintains strong adhesion and structural integrity even after prolonged exposure to sweat and mechanical stress, making it suitable for long-term wearable applications. Furthermore, in vitro biocompatibility tests confirm its potential for biomedical applications. By integrating this hydrogel into a multi-channel sensor system, the study successfully achieves real-time cervical motion monitoring and classification through a CNN-based model, validating its applicability in health monitoring and injury prevention. Overall, this research presents a DLP 3D-printed hydrogel platform with customizable mechanical properties and advanced sensing capabilities, paving the way for future applications in flexible electronics, smart wearable devices, and biomedical sensors. Future work will focus on further optimizing material formulation, scalability, and integration with wireless sensor technologies, expanding its functionality for real-time health monitoring and motion tracking applications.

**Supporting Information**

- Muscle strain sensors. (Movie S1) (**.MP4**)

- Kinesiology Tape - Neck status monitoring. (Movie S2) (**.MP4**)

- Porosity measurement method;…. (**.Word**)

**Acknowledgments:**

The authors thank Takeshi Mino Researcher, Atsuko Nagataki Researcher and Kagami Memorial Research Institute for Materials Science and Technology, Waseda University for helping to experiments. This work was the result of using research equipment (G1023) shared in MEXT Project for promoting public utilization of advanced research infrastructure (Program for supporting construction of core facilities) Grant Number JPMXS0440500022 and JPMXS0440500023; This work was partly executed under the cooperation of organization between Waseda University and Kioxia Corporation, Grant Number B2R50Z003000. This work was supported by JSPS KAKENHI Grant Number JP 23H01382 and the Foundation for Technology Promotion of Electronic Circuit Board. This work was partly executed under the cooperation of organization between Waseda University and Kioxia Corporation,Grant Number B2R50Z003000.

**Credit authorship contribution statement:**

**Ze ZHANG:** Methodology, Writing - review & editing, Visualization, Data curation. **Kewei SONG:** Methodology, Investigation, Writing - review & editing, Visualization. **Kayo HIROSE:** Investigation, Writing - review & editing, Visualization, Supervision, Project administration, Funding acquisition. **Jianxian HE:** Visualization, Data curation. **Qianhao LI:** Visualization, Data curation. **Yannan LI:** Visualization, Data curation. **Yifan PAN:** Visualization, Data curation. **Rongyi ZHUANG:** Data curation. **Mohamed Adel:** Visualization. **Shogo IWAI:** Visualization. **Ahmed M. R. Fath El-Bab:** Investigation, Writing - review & editing, Visualization. **Zhouyuan YANG:** Investigation, Writing - review & editing, Visualization, Funding acquisition. **Hui FANG:** Investigation, Writing - review & editing, Visualization, Funding acquisition. **Shinjiro UMEZU:** Conceptualization, Methodology, Investigation, Writing - review & editing, Visualization, Supervision, Project administration, Funding acquisition.

**Conflict of Interest:**

The authors declare no conflict of interest.

**Data Availability Statement:**

The authors declare that the data supporting the findings of this study are available within the paper and its supplementary information files.

**Ethics approval obtained:**
The methodology for this study was approved by the Human Research Ethics committee of WASEDA University (Ethics approval number: 2024-518).